\documentclass[aps,prd,twocolumn,superscriptaddress,amsmath,amssymb,nofootinbib,floatfix]{revtex4-2}

\usepackage{graphicx}
\usepackage{bm}
\usepackage[colorlinks=true,linkcolor=blue,citecolor=blue,urlcolor=blue]{hyperref}

\newcommand{\re}{r_{e}}
\newcommand{\dd}{\mathrm{d}}

\begin{document}

\title{Lorentz and Poincar\'e holonomies of an acoustic vortex:\\
ergoregion transition, Sagnac time, and beyond-Aharonov--Bohm phonon scattering}

\author{Claudio Furtado}
\email{furtado@fisica.ufpb.br}
\thanks{ORCID: \href{https://orcid.org/0000-0002-3455-4285}{0000-0002-3455-4285}}
\affiliation{Departamento de F\'{\i}sica, CCEN, Universidade Federal da Para\'{\i}ba,
Cidade Universit\'aria, 58051-970 Jo\~ao Pessoa, PB, Brazil}

\author{A.~M.~de~M.~Carvalho}
\email{alexandre@fis.ufal.br}
\thanks{ORCID: \href{https://orcid.org/0009-0006-3540-0364}{0009-0006-3540-0364}}
\affiliation{Instituto de F\'{\i}sica, Universidade Federal de Alagoas,
Campus A.~C. Sim\~oes, 57072-970 Macei\'o, AL, Brazil}

\author{L.~C.~Garcia~de~Andrade}
\affiliation{Departamento de F\'{\i}sica Te\'orica, Instituto de F\'{\i}sica,
Universidade do Estado do Rio de Janeiro, 20550-013 Rio de Janeiro, RJ, Brazil}

\author{Fernando Moraes}
\email{fernando.jsmoraes@ufrpe.br}
\thanks{ORCID: \href{https://orcid.org/0000-0001-7045-054X}{0000-0001-7045-054X}}
\affiliation{Departamento de F\'{\i}sica,
Universidade Federal Rural de Pernambuco,
52171-900 Recife, PE, Brazil}

\date{\today}

\begin{abstract}
Far from its core, a vortex in a superfluid is acoustically equivalent to a
massless spinning cosmic string, and the scattering of phonons reduces to an
Aharonov--Bohm (AB) problem. Beyond this limit, we compute the Lorentz and Poincaré holonomies in closed form, the Lorentz and Poincar\'e holonomies of the \emph{exact} acoustic
geometry of an irrotational vortex for circular loops of arbitrary radius~$r$.
Using the Painlev\'e--Gullstrand tetrad, which is regular inside the
ergoregion, we show that the Lorentz holonomy changes conjugacy class at the
ergosurface $r=\re=\Gamma/2\pi c$: it is a rotation by
$2\pi\sqrt{1-\re^{2}/r^{2}}$ outside, a null rotation on, and a boost of
rapidity $2\pi\sqrt{\re^{2}/r^{2}-1}$ inside. The invariant translational component
of the Poincar\'e holonomy is a time shift
$\tau(r)=(\Gamma/c^{2})(1-v_s^{2}/c^{2})^{-1/2}$, which we identify with half
the acoustic Sagnac delay measured in the proper time of a static observer; it
reduces to the spinning-string value $\Gamma/c^{2}$ as $r\to\infty$. No
torsion is required: the translational holonomy follows from the Cartan
development of the Levi-Civita connection. The non-topological part of the
geometry leaves a long-range imprint on phonon scattering: the centrifugal
term is $m^{2}+2m\alpha$ instead of the AB form $(m+\alpha)^{2}$, with
$\alpha=\omega\Gamma/2\pi c^{2}$, so the $s$ wave is not phase shifted. As a
consequence the transverse cross section, which controls the Iordanskii force,
is $q\sigma_{\perp}=2\pi\alpha-\pi^{3}\alpha^{3}/3+O(\alpha^{5})$, rather
than the AB value $\sin 2\pi\alpha$. We show that the near-zone curvature and
any short-range, parity-symmetric core physics enter $\sigma_\perp$ only at
$O(\alpha^{5})$, so the $\alpha^{3}$ term is universal. We give estimates for
superfluid $^4$He, atomic condensates, and shallow-water analogues.
\end{abstract}

\maketitle

\section{Introduction}
\label{sec:intro}

Analogue gravity exploits the fact that linear perturbations of a moving
fluid propagate as fields on an effective Lorentzian geometry, the
\emph{acoustic metric}~\cite{Unruh1981,Visser1998,BLV2011}. Because the
underlying microphysics is known and accessible, such systems are
laboratories for kinematical aspects of field theory in curved space-time
that are difficult or impossible to probe astrophysically. Following early
work on superfluid helium~\cite{Volovik1998,VolovikBook}, the program has
reached the experimental stage: rotational superradiance was observed in a
draining water vortex~\cite{Torres2017}, and spectroscopy of bound states and
ringdown-like modes has been performed on a giant quantum vortex in
superfluid $^4$He carrying of order $10^{4}$ circulation
quanta~\cite{Svancara2024}.

A quantized vortex is the simplest rotating acoustic geometry. Far from its
core, where the superfluid velocity $v_s=\Gamma/2\pi r$ is small compared to
the speed of sound $c$, the acoustic line element reduces to that of a
massless spinning cosmic string~\cite{Volovik1998,Mazur1986,DJtH1984}, a
locally flat space-time with a helical time structure. Phonons scattered by
the vortex then experience an analogue of the gravitational Aharonov--Bohm
(AB) effect~\cite{AB1959,FordVilenkin1981,Bezerra1987}, with AB parameter
$\alpha=\omega\Gamma/2\pi c^{2}$. This analogy is at the heart of the
long-standing discussion of the Iordanskii force on a moving
vortex~\cite{Iordanskii1966,Sonin1997,Stone2000,TAN1996,Sonin2001}, and it
has been explored in shallow-water experiments that image the dislocated
wavefronts produced by vertical vorticity~\cite{Coste1999,Vivanco1999}.

The exact vortex geometry is, however, not flat. Fischer and
Visser~\cite{FischerVisser2002,FischerVisser2003} showed that the curvature
of the acoustic vortex metric is appreciable at intermediate distances, and
the metric possesses an ergosurface at $r=\re\equiv\Gamma/2\pi c$, where
$v_s=c$. We address two questions: what is the global geometric content of the exact vortex geometry, and how does its non-topological part affect phonon scattering?  Holonomies are the natural tool. For a cosmic string the Lorentz
(rotational) holonomy of a loop encircling the defect encodes the conical
deficit, while for a spinning string the Lorentz holonomy is trivial and the
information sits in the translational part of the Poincar\'e
holonomy~\cite{Bezerra1987,Bezerra1990,Tod1994,Anandan1994}. A preliminary
attempt to use this language for the vortex was made in an unpublished
preprint by the present authors~\cite{Furtado2004}. That work relied on a
non-integrable ``coordinate transformation'', interpreted the result in an
Einstein--Cartan setting with torsion, and did not compute the holonomy of
the curved region. The present paper replaces it with a self-contained and
corrected treatment.

Our results are as follows. (i)~We compute the Lorentz holonomy of the exact
acoustic vortex for circular loops of any radius, in closed form, using the
Painlev\'e--Gullstrand tetrad, and show that its conjugacy class changes from
elliptic to hyperbolic precisely at the ergosurface. (ii)~We compute the
Poincar\'e holonomy and identify its invariant translational part with a
physically measurable quantity, the acoustic Sagnac delay. (iii)~We show
that the departure of the exact geometry from the spinning string modifies
the phonon partial waves at $O(\alpha^{2})$ in a way that is long-ranged and
universal, and compute the resulting correction to the transverse cross
section. In particular, the frequently used replacement of the AB phase
$\Phi$ by $\sin\Phi$ in the Iordanskii force~\cite{Stone2000,Sonin2001} is
not correct beyond linear order for a vortex.

The paper is organized as follows. Section~\ref{sec:metric} reviews the
acoustic vortex geometry and its curvature. Sections~\ref{sec:lorentz}
and~\ref{sec:poincare} contain the holonomy calculations. Phonon scattering
is treated in Sec.~\ref{sec:scattering}, and physical estimates are given in
Sec.~\ref{sec:estimates}. Section~\ref{sec:conclusions} summarizes. All
symbolic results were checked with computer algebra; the scripts are
provided as Supplemental Material.

\section{Acoustic geometry of a vortex}
\label{sec:metric}

Consider a barotropic, inviscid fluid with background velocity
$\mathbf{v}=\nabla\Phi_v$ and small irrotational perturbations
$\delta\mathbf{v}=\nabla\psi$. For constant background density $\rho$ and
sound speed $c$, the perturbation obeys~\cite{Unruh1981,Visser1998}
\begin{equation}
\left(\partial_t+\mathbf{v}\cdot\nabla\right)^{2}\psi
- c^{2}\nabla^{2}\psi=0,
\label{eq:hydro}
\end{equation}
which is the d'Alembert equation
$\Box\psi=(-g)^{-1/2}\partial_\mu\big[(-g)^{1/2}g^{\mu\nu}\partial_\nu\psi\big]=0$
for the acoustic metric
\begin{equation}
\dd s^{2}=-c^{2}\dd t^{2}+\big(\dd\mathbf{x}-\mathbf{v}\,\dd t\big)^{2},
\label{eq:PGgeneral}
\end{equation}
up to an irrelevant constant conformal factor. For a straight vortex along
$z$ with circulation $\Gamma=N\kappa$ ($\kappa=h/m$ for a superfluid of
atoms of mass $m$), $\mathbf{v}=v_s(r)\,\hat{\bm\phi}$ with
$v_s=\Gamma/2\pi r$, and Eq.~\eqref{eq:PGgeneral} becomes
\begin{equation}
\dd s^{2}=-\left(c^{2}-v_s^{2}\right)\dd t^{2}
-\frac{\Gamma}{\pi}\,\dd t\,\dd\phi+\dd r^{2}+r^{2}\dd\phi^{2}+\dd z^{2}.
\label{eq:metric}
\end{equation}
Note that $g_{t\phi}=-\Gamma/2\pi$ is \emph{constant}. We have verified
explicitly that $\Box\psi=0$ for Eq.~\eqref{eq:metric} reproduces
Eq.~\eqref{eq:hydro}, with $\sqrt{-g}=cr$. The $z$ direction is trivial and
is suppressed below. We neglect the depletion of the density near the core
predicted by Bernoulli's law, $\delta\rho/\rho\simeq -v_s^{2}/2c^{2}$; its
effect is discussed in Sec.~\ref{sec:conclusions}.

The metric has a single length scale,
\begin{equation}
\re=\frac{\Gamma}{2\pi c},\qquad \varepsilon(r)\equiv\frac{v_s}{c}=\frac{\re}{r},
\label{eq:re}
\end{equation}
the radius of the ergosurface, where $g_{tt}=0$. For $r<\re$ no observer can
remain at rest with respect to the laboratory.

The massless spinning cosmic string~\cite{Mazur1986,DJtH1984} has line
element $\dd s^{2}_{\rm str}=-c^{2}(\dd t+\beta\,\dd\phi)^{2}+\dd r^{2}
+r^{2}\dd\phi^{2}$, which is flat (set $T=t+\beta\phi$). Choosing
$\beta=\Gamma/2\pi c^{2}$ so that the $\dd t\,\dd\phi$ terms coincide, one
finds the exact relation
\begin{equation}
\dd s^{2}=\dd s^{2}_{\rm str}
+\frac{\re^{2}}{r^{2}}\,c^{2}\dd t^{2}+\re^{2}\,\dd\phi^{2}.
\label{eq:difference}
\end{equation}
The vortex is the spinning string plus a perturbation of relative order
$\re^{2}/r^{2}$. This term carries all the curvature and, as we show in
Sec.~\ref{sec:scattering}, all the departures from AB scattering.

We use the Painlev\'e--Gullstrand (PG) coframe
\begin{equation}
e^{\hat 0}=c\,\dd t,\qquad e^{\hat 1}=\dd r,\qquad
e^{\hat 2}=r\,\dd\phi-v_s\,\dd t,
\label{eq:PGtetrad}
\end{equation}
with $\eta_{ab}={\rm diag}(-1,1,1)$. It is adapted to observers comoving
with the fluid and, unlike the static coframe used in
Refs.~\cite{Volovik1998,Furtado2004}, is regular at and inside the
ergosurface. The torsion-free connection one-forms, defined by
$\dd e^{a}+\omega^{a}{}_{b}\wedge e^{b}=0$, are
\begin{align}
\omega^{\hat 0}{}_{\hat 1}&=\frac{v_s^{2}}{c\,r}\,\dd t-\varepsilon\,\dd\phi,
\qquad
\omega^{\hat 0}{}_{\hat 2}=-\frac{\varepsilon}{r}\,\dd r,\nonumber\\
\omega^{\hat 1}{}_{\hat 2}&=-\dd\phi .
\label{eq:connection}
\end{align}
The curvature is nonzero and falls off as $r^{-4}$. Its invariants are
\begin{equation}
\mathcal{R}=\frac{2\re^{2}}{r^{4}},\qquad
R_{abcd}R^{abcd}=\frac{44\,\re^{4}}{r^{8}},
\label{eq:invariants}
\end{equation}
and the only frame component needed below is
\begin{equation}
R^{\hat 1}{}_{\hat 2}\big|_{r\phi}=-\frac{\re^{2}}{r^{3}} .
\label{eq:R12}
\end{equation}
The full set of components is listed in Appendix~\ref{app:curv}. The
curvature is quadratic in $\Gamma$, in agreement with
Eq.~\eqref{eq:difference}.

\begin{figure*}[t]
\centering
\includegraphics[width=\textwidth]{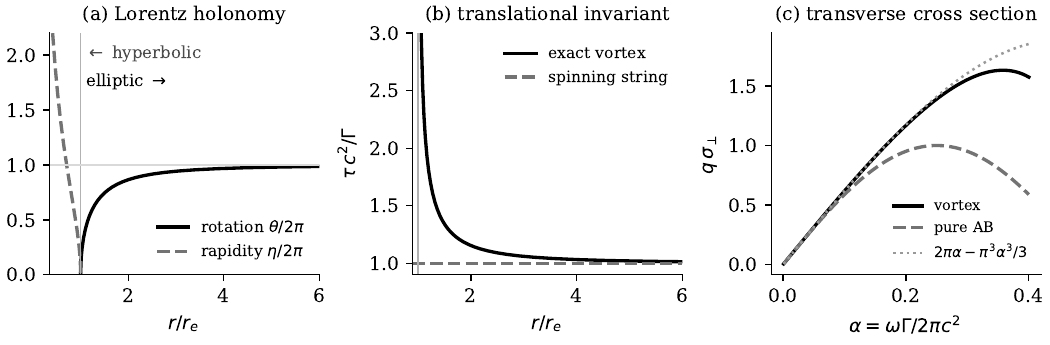}
\caption{(a)~Conjugacy class of the Lorentz holonomy of a circle of radius
$r$ around the vortex: rotation angle $\theta/2\pi$ [Eq.~\eqref{eq:theta}]
for $r>\re$ and boost rapidity $\eta/2\pi$ [Eq.~\eqref{eq:eta}] for $r<\re$.
(b)~Invariant translational holonomy $\tau(r)$, Eq.~\eqref{eq:tau}, compared
with the spinning-string value $\Gamma/c^{2}$. (c)~Transverse cross section
from the exact partial-wave sum with $\nu_m=\sqrt{m^{2}+2m\alpha}$ (solid),
the pure AB result $\sin2\pi\alpha$ (dashed), and the expansion
\eqref{eq:sigmaexp} (dotted).}
\label{fig:holonomy}
\end{figure*}

\section{Lorentz holonomy}
\label{sec:lorentz}

The Lorentz holonomy of a closed curve $\gamma$ is the path-ordered
exponential
\begin{equation}
U(\gamma)=\mathcal{P}\exp\left(-\oint_\gamma\omega\right)\in SO(2,1),
\label{eq:Udef}
\end{equation}
which maps a vector at the base point to its parallel-transported image. A
change of frame conjugates $U$; its conjugacy class is frame independent.

We take $\gamma_r$ to be the circle of radius $r$ at fixed $t$ (and $z$),
traversed counterclockwise. The curve is spacelike for all $r$, since
$g_{\phi\phi}=r^{2}>0$. Along $\gamma_r$ only $\omega_\phi$ contributes, and
in the PG frame it is independent of $\phi$. The path ordering is therefore
trivial and the holonomy is exactly
\begin{equation}
U(r)=\exp\left(-2\pi\mathcal{M}\right),\qquad
\mathcal{M}=\begin{pmatrix}0&-\varepsilon&0\\-\varepsilon&0&-1\\0&1&0\end{pmatrix},
\label{eq:Uexact}
\end{equation}
where $\mathcal{M}=(\omega_\phi)^{a}{}_{b}$. In terms of the generators
$J$ (rotation in the $\hat1\hat2$ plane) and $K$ (boost along $\hat 1$),
$\mathcal{M}=-J-\varepsilon K$. The rotation is the usual one of polar
coordinates; the boost of strength $\varepsilon=v_s/c$ reflects the motion of
the PG frames relative to the laboratory.

Since $\mathcal{M}^{3}=-\lambda^{2}\mathcal{M}$ with
$\lambda^{2}=1-\varepsilon^{2}$, the exponential can be summed in closed form,
\begin{equation}
U(r)=\mathbb{1}-\frac{\sin 2\pi\lambda}{\lambda}\,\mathcal{M}
+\frac{1-\cos 2\pi\lambda}{\lambda^{2}}\,\mathcal{M}^{2},
\label{eq:Uclosed}
\end{equation}
with
\begin{equation}
{\rm tr}\,U(r)=1+2\cos\left(2\pi\sqrt{1-\re^{2}/r^{2}}\right).
\label{eq:trace}
\end{equation}
The Casimir of $\mathcal{M}$ is $J^{2}-K^{2}\propto 1-\varepsilon^{2}$, which
gives the classification:
\begin{itemize}
\item $r>\re$ (\emph{elliptic}): $U$ is conjugate to a spatial rotation by
\begin{equation}
\theta(r)=2\pi\sqrt{1-\re^{2}/r^{2}} .
\label{eq:theta}
\end{equation}
\item $r=\re$ (\emph{parabolic}): $\lambda=0$ and
$U=\mathbb{1}-2\pi\mathcal{M}+2\pi^{2}\mathcal{M}^{2}$ is a null rotation.
\item $r<\re$ (\emph{hyperbolic}): $U$ is conjugate to a boost of rapidity
\begin{equation}
\eta(r)=2\pi\sqrt{\re^{2}/r^{2}-1} .
\label{eq:eta}
\end{equation}
\end{itemize}
These invariants are plotted in Fig.~\ref{fig:holonomy}(a). As a check of
frame independence, we computed the holonomy in the static coframe of
Refs.~\cite{Volovik1998,Furtado2004} (valid only for $r>\re$) and obtained the
same trace~\eqref{eq:trace}.

Three remarks are in order. First, as $r\to\infty$, $\theta\to 2\pi$ and the
holonomy becomes trivial. There is no conical deficit, in agreement with the
massless spinning string, and contrary to the ``conical defect'' language of
Refs.~\cite{Volovik1998,Furtado2004}. Second, at large $r$ the angle deficit
is $2\pi-\theta\simeq\pi\re^{2}/r^{2}$. This is exactly the curvature flux
through the exterior of the loop, computed from Eq.~\eqref{eq:R12},
\begin{equation}
\int_{r}^{\infty}\!\!\dd r'\!\int_{0}^{2\pi}\!\!\dd\phi\;\frac{\re^{2}}{r'^{3}}
=\frac{\pi\re^{2}}{r^{2}},
\label{eq:GB}
\end{equation}
which is a Gauss--Bonnet statement at leading order; at higher orders the
non-Abelian character of $SO(2,1)$ matters. The holonomy is therefore not
topological: it depends on $r$ and probes the curvature outside the
loop~\cite{Vickers1987}. Third, the change of class at $r=\re$ is a
holonomic signature of the ergosurface. Inside it the frame transported
around the vortex returns boosted, which is consistent with the absence of
static observers there.

\section{Poincar\'e holonomy and Sagnac time}
\label{sec:poincare}

For a spinning string the Lorentz holonomy is trivial, and the geometry is
detected only by the translational part of the Poincar\'e
holonomy~\cite{Bezerra1990,Tod1994,Anandan1994}. This object is the Cartan
development of the loop: one rolls the manifold along $\gamma$ on its
tangent Minkowski space and records the affine map relating the initial and
final tangent spaces. It is built from the affine connection
\begin{align}
\mathcal{A}&=\begin{pmatrix}\omega^{a}{}_{b}& e^{a}\\0&0\end{pmatrix},\nonumber\\
H(\gamma)&=\mathcal{P}\exp\left(-\oint_\gamma\mathcal{A}\right)
=\begin{pmatrix}U&\bm a\\0&1\end{pmatrix},
\label{eq:affine}
\end{align}
where the coframe enters as the translational part of the
connection~\cite{Bezerra1990,Tod1994}. Here $\omega$ here is the
\emph{Levi-Civita} connection. No torsion is involved, and none is needed to
obtain a nontrivial translational holonomy. Earlier interpretations of
$\oint e^{\hat 0}$ as a ``torsion flux'' of the
vortex~\cite{Furtado2004} conflated the anholonomy of a
particular coframe with torsion; in the present setting that quantity is
coframe dependent and has no invariant meaning.

For $\gamma_r$ in the PG frame, $\mathcal{A}_\phi$ is again constant, with
$e_\phi=(0,0,r)^{T}$, so that
\begin{equation}
\bm a=-\left(\int_0^{2\pi}\!e^{-s\mathcal{M}}\dd s\right)e_\phi .
\label{eq:a}
\end{equation}
Under a change of base point the translation $\bm a$ changes by
$(\mathbb{1}-U)\bm b$. Hence only its component along the axis $n$ of $U$
($\mathcal{M}n=0$) is invariant. For $r>\re$ the axis is timelike,
\begin{equation}
n=\frac{(1,0,-\varepsilon)}{\sqrt{1-\varepsilon^{2}}} ,
\label{eq:axis}
\end{equation}
and, since $n_a\mathcal{M}^{a}{}_{b}=0$, the invariant follows directly from
Eq.~\eqref{eq:a}: $n\cdot\bm a=-2\pi\,n\cdot e_\phi
=2\pi\varepsilon r/\sqrt{1-\varepsilon^{2}}$. In time units, this gives
\begin{equation}
;\tau(r)=\frac{n\cdot\bm a}{c}
=\frac{\Gamma}{c^{2}}\,\frac{1}{\sqrt{1-v_s^{2}/c^{2}}}
\label{eq:tau}
\end{equation}
(the sign is fixed by the sense of circulation). Figure~\ref{fig:holonomy}(b)
shows $\tau(r)$.
(a)~For the spinning string, the same construction gives $U=\mathbb 1$ and
$\bm a=(2\pi\beta c,0,0)$, i.e.\ a pure time translation $2\pi\beta=\Gamma/c^{2}$.
This agrees with Refs.~\cite{Bezerra1990,Mazur1986}, and Eq.~\eqref{eq:tau}
tends to it as $r\to\infty$.
(b)~The axis~\eqref{eq:axis} has spatial velocity $-v_s$ relative to the PG
(fluid) frame, so it is the four-velocity of a \emph{static} laboratory
observer. Equation~\eqref{eq:tau} is a time interval in that observer's
rest frame.
(c)~This interval is measurable. Sound signals constrained to circulate at
radius $r$ (e.g.\ in an annular waveguide) take coordinate times
$2\pi r/(c\pm v_s)$ in the two senses, so the Sagnac delay is
$\Delta t=2\Gamma/[c^{2}(1-v_s^{2}/c^{2})]$. A static observer measures
proper time $\dd\tau_{\rm st}=\sqrt{1-v_s^{2}/c^{2}}\;\dd t$, and therefore
\begin{equation}
\Delta\tau_{\rm st}=2\,\tau(r).
\label{eq:sagnac}
\end{equation}
The invariant translational holonomy is half the acoustic Sagnac delay
measured by a static clock. In this sense the ``gravitational AB effect'' of
the vortex is a Sagnac effect, a connection long recognized for rotating
frames, now made precise holonomically at any radius outside the
ergosurface.
(d)~For $r<\re$ the axis is spacelike. The invariant
$2\pi\re/\sqrt{\re^{2}/r^{2}-1}$ (in length units) is then the pitch of a spatial
screw motion, and no clock interpretation exists, as expected inside an
ergoregion.

The function $\beta(r)=\Gamma/[2\pi c^{2}(1-v_s^{2}/c^{2})^{1/2}]$ appeared
in Ref.~\cite{Furtado2004} as a coefficient of a non-integrable coframe. The
present analysis shows that $2\pi\beta(r)$ is in fact the gauge-invariant
translational holonomy, and gives it a direct operational meaning.

\section{Phonon scattering beyond the Aharonov--Bohm limit}
\label{sec:scattering}

\subsection{Partial waves}

With $\psi=e^{-i\omega t+im\phi+ik_z z}\,R(r)$, Eq.~\eqref{eq:hydro} for the
metric~\eqref{eq:metric} gives, exactly,
\begin{equation}
R''+\frac{R'}{r}+\left[q^{2}-\frac{m^{2}+2m\alpha}{r^{2}}
+\frac{\re^{2}m^{2}}{r^{4}}\right]R=0,
\label{eq:radial}
\end{equation}
where $q^{2}=\omega^{2}/c^{2}-k_z^{2}$ and
\begin{equation}
\alpha=\frac{\omega\Gamma}{2\pi c^{2}}=\frac{\omega}{c}\,\re .
\label{eq:alpha}
\end{equation}
For the spinning string, or equivalently for the minimally coupled equation
$c^{2}(\nabla+i\omega\mathbf{v}/c^{2})^{2}\psi=-\omega^{2}\psi$ used in
Ref.~\cite{Stone2000}, the bracket is instead $q^{2}-(m+\alpha)^{2}/r^{2}$,
which is the AB problem with flux parameter $\alpha$. The two differ by
\begin{equation}
\frac{(m+\alpha)^{2}-(m^{2}+2m\alpha)}{r^{2}}=\frac{\alpha^{2}}{r^{2}},
\qquad\text{and}\qquad \frac{\re^{2}m^{2}}{r^{4}} .
\label{eq:diff}
\end{equation}
These are precisely the contributions of the two terms in
Eq.~\eqref{eq:difference}. The first is long ranged and independent of $m$;
the second is short ranged and attractive, and it dominates only for
$r\lesssim\re$, inside the region where the hydrodynamic description fails
for a quantized vortex (Sec.~\ref{sec:estimates}). The index
$\sqrt{m^{2}+2m\alpha}$ was obtained for the draining-bathtub vortex in the
asymptotic region in Ref.~\cite{DolanOliveira2013}.

Treating the inverse-square part exactly, the regular solutions are
$J_{\nu_m}(qr)$ with
\begin{equation}
\nu_m=\sqrt{m^{2}+2m\alpha},\qquad
\delta^{(0)}_m=\frac{\pi}{2}\big(|m|-\nu_m\big),
\label{eq:nu}
\end{equation}
to be compared with $\nu_m^{\rm AB}=|m+\alpha|$. For $|m|\geq1$,
\begin{equation}
\nu_{\pm|m|}=|m|\pm\alpha-\frac{\alpha^{2}}{2|m|}\pm\frac{\alpha^{3}}{2m^{2}}+O(\alpha^{4}),
\label{eq:nuexp}
\end{equation}
so the parity-odd (AB-like) part of the phase shifts,
$\nu_{|m|}-\nu_{-|m|}=2\alpha+\alpha^{3}/m^{2}+\dots$, coincides with AB up
to $O(\alpha^{3})$. The parity-even part acquires a correction
$-\alpha^{2}/|m|$. The essential difference is in the $s$ wave. For $m=0$ the
vortex gives $\nu_0=0$, whereas AB gives $\nu_0^{\rm AB}=\alpha$. A vortex
does not couple to an axisymmetric wave, since the coupling in
Eq.~\eqref{eq:hydro} is $\mathbf{v}\cdot\nabla\propto\partial_\phi$, and so
it does not shift the $s$ wave. An AB flux does shift it.

The $r^{-4}$ term adds, in Born approximation on the Bessel basis,
\begin{equation}
\delta^{(4)}_m=\frac{\pi}{2}\,\re^{2}m^{2}q^{2}\!\int_0^\infty\!\frac{J_{|m|}^{2}(x)}{x^{3}}\dd x
=\frac{\pi\alpha^{2}}{8}\,\frac{|m|}{m^{2}-1},
\label{eq:born}
\end{equation}
for $|m|\geq2$ and $k_z=0$ (Appendix~\ref{app:born}). The $|m|=1$ integral
is logarithmically sensitive to the core. Both contributions are parity
even. Adding Eqs.~\eqref{eq:nu} and \eqref{eq:born} at large $|m|$ gives
\begin{equation}
\delta_m\simeq-\frac{\pi\alpha}{2}\,{\rm sgn}\,m+\frac{3\pi\alpha^{2}}{8|m|},
\label{eq:dolan}
\end{equation}
in agreement with the Born result of Dolan and Oliveira~\cite{DolanOliveira2013}
in the non-draining limit. This is an independent check of
Eqs.~\eqref{eq:radial}--\eqref{eq:born}.

\subsection{Transverse cross section and the Iordanskii force}

The transverse (Magnus-like) cross section, which determines the Iordanskii
force exerted by a phonon gas on the vortex, is~\cite{Sonin1997,Sonin2001} (up to an overall sign convention)
\begin{equation}
\sigma_\perp=\frac{1}{q}\sum_{m=-\infty}^{\infty}\sin 2\left(\delta_m-\delta_{m+1}\right).
\label{eq:sigmaperp}
\end{equation}
For AB phase shifts it gives $q\sigma_\perp^{\rm AB}=\sin2\pi\alpha$. This
result motivated the replacement $\Phi\to\sin\Phi$ in the Iordanskii
force~\cite{Stone2000,Sonin2001}.

With the vortex phase shifts~\eqref{eq:nu}, the sum is dominated at small
$\alpha$ by the two terms adjacent to the unshifted $s$ wave:
$\delta_{-1}-\delta_0\simeq\delta_0-\delta_1\simeq\pi\alpha/2$, instead of
$\pi\alpha$ and $0$ for AB. Including the $O(\alpha^{2}/|m|)$ terms of the
other partial waves, which contribute only at higher order, we find
\begin{equation}
q\,\sigma_\perp=2\pi\alpha-\frac{\pi^{3}}{3}\alpha^{3}+O(\alpha^{5}),
\label{eq:sigmaexp}
\end{equation}
to be compared with $q\sigma^{\rm AB}_\perp=2\pi\alpha-\tfrac{4\pi^{3}}{3}\alpha^{3}+O(\alpha^{5})$, i.e.
\begin{equation}
q\left(\sigma_\perp-\sigma^{\rm AB}_\perp\right)=\pi^{3}\alpha^{3}+O(\alpha^{5}).
\label{eq:sigmadiff}
\end{equation}
We verified Eqs.~\eqref{eq:sigmaexp} and \eqref{eq:sigmadiff} by summing
Eq.~\eqref{eq:sigmaperp} numerically with $|m|\le 2\times10^{4}$. The
coefficient of $\alpha^{3}$ in Eq.~\eqref{eq:sigmadiff} converges to
$31.00\simeq\pi^{3}$ [Fig.~\ref{fig:holonomy}(c)]. The leading term, and
hence the Iordanskii force at linear order in $\Gamma$, is unchanged. The
first correction is suppressed four times less than the AB formula predicts:
$q\sigma_\perp/2\pi\alpha=1-\pi^{2}\alpha^{2}/6$ instead of
$1-2\pi^{2}\alpha^{2}/3$.

Equation~\eqref{eq:sigmaexp} would be of limited value if it depended on the
poorly known near-zone and core physics. It does not. Let $s_{|m|}$ be any
parity-even addition to the phase shifts, such as the $r^{-4}$
term~\eqref{eq:born}, a soft core, or a density depletion. To first order,
its contribution to Eq.~\eqref{eq:sigmaperp} is
$\sum_m 2\cos 2\Delta_m^{(0)}\,(s_{|m|}-s_{|m+1|})$, with
$\Delta_m^{(0)}=\delta^{(0)}_m-\delta^{(0)}_{m+1}$. The part with
$\cos\to1$ telescopes to zero. The $O(\alpha^{2})$ part of the cosine is
nonzero only for $m=0,-1$, where
$\Delta^{(0)}_0=\Delta^{(0)}_{-1}+O(\alpha^{2})$, and the two terms cancel at
this order. The remainder is $O(\alpha^{3}s)+O(\alpha s^{2})$. For
$s=O(\alpha^{2})$, as for Eq.~\eqref{eq:born}, or $s=O(q^{2}a^{2})$ for a
small core of radius $a$, the correction is $O(\alpha^{5})$. Numerically,
adding Eq.~\eqref{eq:born} with an arbitrary core-dependent $s_1$ changes
$q\sigma_\perp$ by an amount scaling as $\alpha^{5}$, as predicted. The
$\alpha^{3}$ term in Eq.~\eqref{eq:sigmaexp} is therefore a universal,
long-distance property of the vortex geometry, fixed by the term
$\re^{2}c^{2}\dd t^{2}/r^{2}$ of Eq.~\eqref{eq:difference}.

The same physics explains the result of Ref.~\cite{DolanOliveira2013} that,
at low frequency, the differential cross section of a non-draining vortex is
$\propto\alpha^{2}\cot^{2}(\varphi/2)$ instead of the AB
$\propto\sin^{2}(\pi\alpha)/\sin^{2}(\varphi/2)$. The two differ by an
isotropic term, which is the missing $s$-wave amplitude.

\subsection{Adiabatic transport}

The holonomy of Sec.~\ref{sec:poincare} also fixes the geometric phase of a
localized phonon wave packet carried adiabatically around the vortex. For
$r\gg\re$ the packet acquires $e^{-i\omega\tau}$ with $\tau\to\Gamma/c^{2}$,
i.e.\ the phase $2\pi\alpha$. This is the result of the non-Abelian Berry
phase analysis for the spinning string~\cite{Mostafazadeh1998,CorichiPierri1995},
which therefore applies to the vortex in this limit. At finite $r$ the
transported packet must be held by an external agent, and its phase depends
on the frames that carry it. Heuristically, for packets held at rest in the laboratory, the
relevant invariant is the static-observer time $\tau(r)$ of
Eq.~\eqref{eq:tau}, measured with the local frequency
$\omega/\sqrt{1-v_s^{2}/c^{2}}$. This gives the phase
$\omega\Gamma/[c^{2}(1-v_s^{2}/c^{2})]=\omega\Delta t/2$, consistent with the
Sagnac interpretation.

\section{Physical estimates}
\label{sec:estimates}

The expansion parameter of Sec.~\ref{sec:scattering} is
$\alpha=\omega\re/c$, the ratio of $\re$ to the reduced wavelength. The
hydrodynamic description requires the wavelength to exceed the core radius
$a$ and the healing (or dispersion) length. Table~\ref{tab:estimates}
collects representative values.

\emph{Superfluid $^4$He, $N=1$.} With $\kappa=h/m_4=9.97\times10^{-8}\,{\rm m^2/s}$
and $c=238$~m/s, we find $\re=0.67$~\AA, which is smaller than the core radius
$a\approx1$~\AA. The ergosurface and the hyperbolic regime of
Fig.~\ref{fig:holonomy}(a) are hidden in the core, and the near-zone
holonomy is of formal interest only. Thermal phonons ($q\sim k_BT/\hbar c$)
have $\alpha\simeq0.011$, $0.018$, $0.037$ at $T=0.3$, $0.5$, $1$~K. The
relative deviation of $\sigma_\perp$ from the AB estimate is
$\pi^{2}\alpha^{2}/2\approx6\times10^{-4}$--$7\times10^{-3}$ in this range. This is small, but it is the leading
non-topological effect and is universal.

\emph{Atomic BEC.} For a Gross--Pitaevskii fluid with healing length
$\xi=\hbar/\sqrt2\,mc$ one finds exactly $\re=\hbar/mc=\sqrt2\,\xi$. The
ergosurface lies at the edge of the core, and $\alpha=\sqrt2\,q\xi$. Here
the $O(\alpha^{2})$ effects compete with Bogoliubov dispersion, which is
also $O(q^{2}\xi^{2})$. A quantitative comparison requires including
dispersion and is left for future work.

\emph{Macroscopic circulation.} In classical fluids and in giant quantum
vortices~\cite{Svancara2024}, $\Gamma$ is not quantized, and $\alpha$ can be
of order $0.1$--$0.3$ at long wavelengths. For shallow-water waves
($c=\sqrt{gh}$) with $h=5$~mm, $\Gamma=5\times10^{-3}\,{\rm m^2/s}$ and
$f=3$~Hz, one finds $\re\approx3.6$~mm and $\alpha\approx0.31$, while
$qh\approx0.43$. This is the regime of
Refs.~\cite{Coste1999,Vivanco1999,Torres2017}. The difference between Eq.~\eqref{eq:sigmaexp} and the AB value is then
large: the exact partial-wave sums give $\sigma_\perp/\sigma_\perp^{\rm AB}\approx1.7$. A quantitative prediction requires
accounting for the residual dispersion, the drain, and the vortex core, but
the universality argument above shows that short-range parity-even effects
do not affect it at $O(\alpha^{3})$.

\begin{table}[!htb]
\caption{Representative scales. $\re=\Gamma/2\pi c$; $\alpha=\omega\re/c$;
$\delta\sigma_\perp/\sigma_\perp$ is the deviation of
Eq.~\eqref{eq:sigmaexp} from the AB value, $\simeq\pi^{2}\alpha^{2}/2$.}
\label{tab:estimates}
\begin{ruledtabular}
\begin{tabular}{lccc}
System & $\re$ & $\alpha$ & $\delta\sigma_\perp/\sigma_\perp$\\
\hline
$^4$He, $N=1$, $T=0.5$ K & $0.67$ \AA & $0.018$ & $1.7\times10^{-3}$\\
$^4$He, $N=1$, $T=1$ K & $0.67$ \AA & $0.037$ & $6.6\times10^{-3}$\\
BEC (GP) & $\sqrt2\,\xi$ & $\sqrt2\,q\xi$ & $\pi^{2}q^{2}\xi^{2}$\\
Shallow water$^{\rm a}$ & $3.6$ mm & $0.31$ & $0.47^{\rm b}$\\
\end{tabular}
\end{ruledtabular}
\begin{flushleft}\footnotesize
$^{\rm a}$\,$h=5$ mm, $\Gamma=5\times10^{-3}$ m$^2$/s, $f=3$ Hz (illustrative).\\
$^{\rm b}$\,Leading-order estimate; the exact partial-wave sums give $0.69$ at $\alpha=0.31$.
\end{flushleft}
\end{table}

\section{Discussion and conclusions}
\label{sec:conclusions}

We have characterized the Lorentz and Poincaré holonomies of the exact acoustic geometry of an irrotational vortex. The Lorentz holonomy of a circle of
radius $r$ is a rotation by $2\pi\sqrt{1-\re^{2}/r^{2}}$ outside the
ergosurface and a boost inside it, and the class change at $r=\re$ provides a
loop-based diagnostic of the ergoregion. The invariant translational
holonomy is the static-observer time $\tau(r)=(\Gamma/c^{2})(1-v_s^{2}/c^{2})^{-1/2}$,
which equals half the acoustic Sagnac delay and reduces to the
spinning-string time shift at large distance. Both results follow from the
torsion-free Levi-Civita connection; the Einstein--Cartan interpretation of
our earlier preprint~\cite{Furtado2004} is unnecessary and is superseded by the present analysis.

On the dynamical side, the non-flat part of the geometry modifies phonon
scattering through a long-range $\alpha^{2}/r^{2}$ term. The vortex does not
shift the $s$ wave, and as a result the transverse cross section is
$q\sigma_\perp=2\pi\alpha-\pi^{3}\alpha^{3}/3+O(\alpha^{5})$ instead of
$\sin2\pi\alpha$. This term is insensitive to the core and to the near-zone
curvature. The Iordanskii force is thus unchanged at linear order in
$\Gamma$, but its saturation at larger $\alpha$ is weaker than the AB-based
estimate.

Several extensions are natural. (i)~Density depletion: Bernoulli's law gives
$\delta\rho/\rho\simeq-\re^{2}/2r^{2}$ and a corresponding change of $c$. In
$3+1$ dimensions this modifies the conformal factor of the acoustic metric
and adds a parity-even $1/r^{2}$ potential, which shifts the even phase
shifts at $O(\alpha^{2})$, including possibly the $s$ wave. Repeating the
universality argument with a parity-even $s_0\neq0$ shows that its effect on
$\sigma_\perp$ is $O(\alpha\,s_0^{2})$, i.e.\ beyond $O(\alpha^{3})$, but an
explicit calculation is desirable. (ii)~Dispersion: for BECs and surface
waves, modified dispersion relations must be included at the same
order~\cite{BLV2011}. (iii)~Draining flows~\cite{DolanOliveira2013,Torres2017}:
the radial velocity adds a horizon, and the Poincar\'e holonomy should then
distinguish horizon and ergosurface. (iv)~Spinning quasiparticles, such as
the fermionic excitations of $^3$He or polarized waves, couple directly to
the Lorentz holonomy computed here, and the elliptic-to-hyperbolic transition
of Fig.~\ref{fig:holonomy}(a) should then have spectroscopic consequences.

\begin{acknowledgments}
C.F. thanks the Conselho Nacional de Desenvolvimento Cient\'{\i}fico e
Tecnol\'ogico (CNPq) for financial support through the research productivity
grant No.~311781/2021-7.
\end{acknowledgments}

\appendix

\section{Connection and curvature in the PG frame}
\label{app:curv}

For the coframe~\eqref{eq:PGtetrad}, the exterior derivatives are
$\dd e^{\hat0}=\dd e^{\hat1}=0$ and, using $v_s'=-v_s/r$,
\begin{equation}
\dd e^{\hat2}=\frac{1}{r}\,e^{\hat1}\wedge e^{\hat2}
+\frac{2\varepsilon}{r}\,e^{\hat1}\wedge e^{\hat0},
\end{equation}
from which Eq.~\eqref{eq:connection} follows. The curvature two-forms
$\Omega^{a}{}_{b}=\dd\omega^{a}{}_{b}+\omega^{a}{}_{c}\wedge\omega^{c}{}_{b}$
have the coordinate components
\begin{align}
\Omega^{\hat0}{}_{\hat1}&=\frac{3\re^{2}c}{r^{4}}\,\dd t\wedge\dd r,\qquad
\Omega^{\hat0}{}_{\hat2}=-\frac{\re^{2}c}{r^{3}}\,\dd t\wedge\dd\phi,\nonumber\\
\Omega^{\hat1}{}_{\hat2}&=-\frac{\re^{3}c}{r^{5}}\,\dd t\wedge\dd r
-\frac{\re^{2}}{r^{3}}\,\dd r\wedge\dd\phi ,
\end{align}
which yield the invariants~\eqref{eq:invariants}. The coefficient of
$\dd r\wedge\dd\phi$ in $\Omega^{\hat1}{}_{\hat2}$ is the component used in
Eq.~\eqref{eq:GB}.

\section{Born phase shift of the $r^{-4}$ term}
\label{app:born}

For a perturbation $\delta V(r)=-\re^{2}m^{2}/r^{4}$ of the radial
operator~\eqref{eq:radial}, the two-dimensional Born phase shift is
$\delta_m=-(\pi/2)\int_0^\infty r\,\delta V\,J_{|m|}^{2}(qr)\,\dd r$. With
$\int_0^\infty J_\mu^{2}(x)\,x^{-3}\dd x=[4\mu(\mu^{2}-1)]^{-1}$ for
$\mu>1$, which we checked numerically, and $q\re=\alpha$ for $k_z=0$,
Eq.~\eqref{eq:born} follows. For $|m|=1$ the integral diverges
logarithmically at small $x$ and must be cut off at the core, $x\sim qa$.
This gives $\delta_{\pm1}\sim(\pi\alpha^{2}/8)\ln(1/qa)$, which is parity even
and therefore covered by the universality argument of
Sec.~\ref{sec:scattering}.



\begin{thebibliography}{99}

\bibitem{Unruh1981}
W.~G. Unruh, Phys. Rev. Lett. \textbf{46}, 1351 (1981).

\bibitem{Visser1998}
M. Visser, Class. Quantum Grav. \textbf{15}, 1767 (1998).

\bibitem{BLV2011}
C. Barcel\'o, S. Liberati, and M. Visser, Living Rev. Relativ. \textbf{14}, 3 (2011).

\bibitem{Volovik1998}
G.~E. Volovik, JETP Lett. \textbf{67}, 881 (1998) [arXiv:cond-mat/9804308].

\bibitem{VolovikBook}
G.~E. Volovik, \emph{The Universe in a Helium Droplet} (Oxford University Press, Oxford, 2003).

\bibitem{Torres2017}
T. Torres, S. Patrick, A. Coutant, M. Richartz, E.~W. Tedford, and S. Weinfurtner,
Nat. Phys. \textbf{13}, 833 (2017).

\bibitem{Svancara2024}
P. \v{S}van\v{c}ara, P. Smaniotto, L. Solidoro, J.~F. MacDonald, S. Patrick,
R. Gregory, C.~F. Barenghi, and S. Weinfurtner, Nature \textbf{628}, 66 (2024).

\bibitem{Mazur1986}
P.~O. Mazur, Phys. Rev. Lett. \textbf{57}, 929 (1986).

\bibitem{DJtH1984}
S. Deser, R. Jackiw, and G. 't Hooft, Ann. Phys. (N.Y.) \textbf{152}, 220 (1984).

\bibitem{AB1959}
Y. Aharonov and D. Bohm, Phys. Rev. \textbf{115}, 485 (1959).

\bibitem{FordVilenkin1981}
L.~H. Ford and A. Vilenkin, J. Phys. A \textbf{14}, 2353 (1981).

\bibitem{Bezerra1987}
V.~B. Bezerra, Phys. Rev. D \textbf{35}, 2031 (1987).

\bibitem{Iordanskii1966}
S.~V. Iordanskii, Zh. Eksp. Teor. Fiz. \textbf{49}, 225 (1965)
[Sov. Phys. JETP \textbf{22}, 160 (1966)].

\bibitem{Sonin1997}
E.~B. Sonin, Phys. Rev. B \textbf{55}, 485 (1997).

\bibitem{Stone2000}
M. Stone, Phys. Rev. B \textbf{61}, 11780 (2000).

\bibitem{TAN1996}
D.~J. Thouless, P. Ao, and Q. Niu, Phys. Rev. Lett. \textbf{76}, 3758 (1996).

\bibitem{Sonin2001}
E.~B. Sonin, arXiv:cond-mat/0104221 (2001).

\bibitem{Coste1999}
C. Coste, F. Lund, and M. Umeki, Phys. Rev. E \textbf{60}, 4908 (1999).

\bibitem{Vivanco1999}
F. Vivanco, F. Melo, C. Coste, and F. Lund, Phys. Rev. Lett. \textbf{83}, 1966 (1999).

\bibitem{FischerVisser2002}
U.~R. Fischer and M. Visser, Phys. Rev. Lett. \textbf{88}, 110201 (2002).

\bibitem{FischerVisser2003}
U.~R. Fischer and M. Visser, Ann. Phys. (N.Y.) \textbf{304}, 22 (2003).

\bibitem{Bezerra1990}
V.~B. Bezerra, Ann. Phys. (N.Y.) \textbf{203}, 392 (1990).

\bibitem{Tod1994}
K.~P. Tod, Class. Quantum Grav. \textbf{11}, 1331 (1994).

\bibitem{Anandan1994}
J. Anandan, Phys. Lett. A \textbf{195}, 284 (1994).

\bibitem{Furtado2004}
C. Furtado, A.~M. de M. Carvalho, L.~C. Garcia de Andrade, and F. Moraes,
arXiv:gr-qc/0401025 (2004).

\bibitem{Vickers1987}
J.~A.~G. Vickers, Class. Quantum Grav. \textbf{4}, 1 (1987).

\bibitem{DolanOliveira2013}
S.~R. Dolan and E.~S. Oliveira, Phys. Rev. D \textbf{87}, 124038 (2013).

\bibitem{Mostafazadeh1998}
A. Mostafazadeh, J. Phys. A \textbf{31}, 7829 (1998).

\bibitem{CorichiPierri1995}
A. Corichi and M. Pierri, Phys. Rev. D \textbf{51}, 5870 (1995).

\end{thebibliography}
\end{document}